\newif\ifhide
\hidetrue

\newif\iftwocolumn
\newif\ifonecolumn
\newif\iflncs
\newif\ifccs
\newif\ifanonymous
\newif\iftoday
\todaytrue
\lncstrue

\ifccs
  \twocolumntrue
\fi
\iflncs
  \onecolumntrue
\fi

\iflncs
\documentclass[11pt]{llncs}
\fi
\ifccs
\documentclass[sigconf, anonymous]{acmart}
\input{ccs-preamble}
\fi

\usepackage{preamble}
\usepackage{multirow}
\usepackage{geometry}
\begin{document}
\title{
  An overview of the efficiency\\and censorship-resistance guarantees\\of widely-used consensus protocols
}
\ifanonymous{\iflncs
\author{}\institute{}
\fi}
\else
\author{
  Orestis Alpos\inst{1}\and
  Bernardo David\inst{1,2}\and
  Nikolas Kamarinakis\inst{1}\and
  Dionysis Zindros\inst{1}
}
\iflncs
\institute{
  Common Prefix\footnote{This work was funded by Flashbots.}\and
  IT University of Copenhagen (ITU)
}
\else
\affiliation{
\institution{
  University\and
  Common Prefix
}
}
\fi
\fi

\iflncs
\maketitle
\fi


\begin{abstract}
Censorship resistance with short-term inclusion guarantees
is an important feature of decentralized systems,
missing from many state-of-the-art and even deployed consensus protocols.
In leader-based protocols the leader arbitrarily selects the transactions to be included in the new block,
and so does a block builder in protocols such as Bitcoin and Ethereum.

In a different line of work, since the redundancy of consensus for implementing distributed payments was formally proven,
consensusless protocols have been described in theory and deployed in the real world.
This has resulted in blockchains and payment systems that are more efficient,
and at the same time avoid the centralized role of a leader or block builder.

In this report we review existing consensus and consensusless protocols with regard to their censorship-resistance, efficiency,
and other properties.
Moreover, we present an approach for new constructions with these properties in mind,
building on existing leader-based protocols.
\end{abstract}

\ifccs
\input{ccs-keywords}
\maketitle
\fi

\section{Model and preliminaries} \label{sec:prelim}

Censorship resistance has been studied in multiple works~\cite{DBLP:conf/ccs/KelkarDLJK23,DBLP:conf/nsdi/YangPAKT22,DBLP:journals/corr/abs-2307-10185}.
In this report we use the following definition of \emph{short-term censorship resistance}.
\begin{definition} [Short-term censorship resistance]
    A protocol is \emph{short-term censorship resistant} if malicious replicas cannot censor honest clients.
    Particularly, they cannot do so even for a short term, meaning that, if the system makes progress and creates an output, then
    a transaction submitted by an honest user appears in it.
\end{definition}

\dotparagraph{Model}
We distinguish two types of parties, $n$ \emph{replicas} (also called \emph{validators} in other works) and an unlimited number of \emph{clients}.
Replicas run the protocol and clients interact with it using its interface.
We call a party -- replica or client -- \emph{honest}, if it follows the protocol, and \emph{malicious} otherwise.
Notice that malicious parties can arbitrarily deviate from protocol specifications.

\dotparagraph{Network}
We review protocols modelled in the synchronous, partially synchronous, and asynchronous model.
In the synchronous model, every message is delivered within a \emph{known} finite delay of $\Delta$ rounds. In the partially synchronous model, messages are delivered within a known finite delay of $\Delta$ rounds only \emph{after} an unknown but finite period of time called the Global Stabilization Time (GST).
Finally, in the asynchronous model, messages are delivered within an \emph{unknown} but finite delay.
We denote by $\delta$ the actual (but unknown) average network delay between two honest replicas.
In synchrony it holds that $\delta\leq \Delta$.
Protocols that proceed in the actual network speed without waiting for $\Delta$-timeouts are called \emph{responsive}.

\dotparagraph{Structure}
In \Cref{sec:previous} we review existing consensus and consensusless protocols with regard to their censorship-resistance, efficiency,
and other properties.
In \Cref{sec:mini-blocks-construction} we present a new approach for new constructions with these properties in mind,
building on existing leader-based protocols.
In \cref{sec:economic-args} we approach censorship-resistance based on rational and incentive-based arguments,
and in \Cref{sec:conclusion} we summarize our findings and make our recommendations.
Finally, \Cref{sec:implementations} gives some links to existing implementations of the protocols we discuss.

\section{Existing protocols}\label{sec:previous}

\subsection{Metrics and comparison result}\label{sec:comparison}
In this section we review and compare existing protocols. \Cref{tab:comparison} reports the following metrics.
\vspace{-1mm}
\begin{itemize}
  \item \emph{Proposal latency} measures the communication rounds from the proposal of a transaction (by one or more replicas) until it becomes committed by $2f+1$ replicas.
  \item \emph{Proposal period} measures the communication rounds between consecutive proposals.
  \item \emph{Max tx censorship} reports the maximum number of rounds the adversary can delay a \emph{specific} transaction, assuming the client sent it to all replicas,
  until it appears in a proposal, and while the rest of the system makes progress.
  \item \emph{Communication complexity} reports the exchanged number of messages in the happy path (column \emph{best}) and in the worst case (column \emph{worst}).
  \item \emph{Throughput} and \emph{latency} report the throughput and latency for WAN deployments under no faults, as reported on the cited papers.
\end{itemize}

\begin{table}[h]
  \caption{Summary and comparison of various protocols.
  For leader-based protocols \emph{max tx censorship} equals the proposal period times $f$, as up to $f$ successive leaders can produce blocks without the censored transaction.
  For DAG-based protocols \emph{max tx censorship} is $0$, because a transaction sent to sufficiently many replicas will be included in the output when the protocol commits the next round.
  \emph{IL} stands for the inclusion lists approach (\Cref{sec:mini-blocks-construction}).
  For these protocols, metrics that start with $+$ indicate an \emph{incremental} additive overhead to a leader-based protocol (the parameters are explained in the corresponding subsection).
  For IL constructions \emph{max tx censorship} equals $0$, as the leader is limited by the ILs and cannot censor transactions.
  We denote by \txsize the transaction size.
  }
  \label{tab:comparison}
  \centering
  \resizebox{\columnwidth}{!}{
  \begin{tabular}{|c|c|c|c|c|c|c|c|c|c|}
    \hline
    \multirow{3}{*}{\textbf{Protocol}} & \multicolumn{5}{c|}{\textbf{Theoretical efficiency}}  & \multicolumn{4}{c|}{\textbf{Bench. (WAN, no faults)}}  \\
    \cline{2-10}
    & \multirow{2}{*}{\begin{tabular}[c]{@{}c@{}}\textbf{Proposal}\\\textbf{latency}\end{tabular}} & \multirow{2}{*}{\begin{tabular}[c]{@{}c@{}}\textbf{Proposal}\\\textbf{period}\end{tabular}}  &  \multirow{2}{*}{\begin{tabular}[c]{@{}c@{}}\textbf{Max tx}\\\textbf{censorship}\end{tabular}}    & \multicolumn{2}{c|}{\textbf{Com. compl.}}  &&\textbf{thr.} &\textbf{lat.}&\\
    \cline{5-6}
    &&  &  &\textbf{best} & \textbf{worst} & \textbf{\#rep} & \textbf{(tps)} & \textbf{(sec)}  & \textbf{ref.}\\
    \hline
    \hline
    Tendermint~\cite{DBLP:journals/corr/abs-1807-04938} & $3\Delta$ & $3\Delta$ & $3\Delta f$ &$O(n^2)$ &$O(n^2)$ & 32& $520$& $2.83$ &\cite[Fig.6]{DBLP:conf/srds/CasonFM0BP21}\\
    &&&&&&100& $500$& 1--9 &\cite{cometBft-qa}\\

    HotStuff/DiemBFT~\cite{DBLP:conf/podc/YinMRGA19} & $7\delta$ & $2\delta$ & $2\Delta f$& $O(n)$& $O(n^2)$ &20& $50K$& $1.7$ &\cite[Fig.5]{DBLP:conf/fc/GelashviliKSSX22}\\

    Jolteon~\cite{DBLP:conf/fc/GelashviliKSSX22} & $5\delta$ & $2\delta$  & $2\Delta f$& $O(n)$& $O(n^2)$ & 20& $50K$& $2$ &\cite[Fig.5]{DBLP:conf/fc/GelashviliKSSX22}\\
    Ditto~\cite{DBLP:conf/fc/GelashviliKSSX22} & $5\delta$ & $2\delta$  & $2\Delta f$& $O(n)$& $O(n^2)$ & 20& $50K$& $1.5$ &\cite[Fig.5]{DBLP:conf/fc/GelashviliKSSX22}\\

    HotStuff-2~\cite{DBLP:journals/iacr/MalkhiN23} & $5\delta$ & $2\delta$  & $2\Delta f$& $O(n)$& $O(n^2)$ & \multicolumn{4}{c|}{}\\

    MoonShot\cite{DBLP:journals/corr/abs-2401-01791} & $5\delta$ & $\delta$  & $\Delta f$& $O(n^2)$& $O(n^2)$ & \multicolumn{4}{c|}{}\\

     \hline

     IL ~(Sec.\ref{sec:inclusion-lists-base})    &$+0$ & $+0$  &$0$    &\multicolumn{2}{c|}{$+O(n^2 |\inclist|)$} & \multicolumn{4}{c|}{}    \\

     IL w. DA~(Sec.\ref{sec:incl-list-da})    &$+\dadisp + \daretr$ & $+0$  &$0$    &\multicolumn{2}{c|}{$+\dacom$} & \multicolumn{4}{c|}{}     \\

     IL w. b/cast~(Sec.\ref{sec:incl-list-bcast})    &$+2\delta$ & $+0$  &$0$   &\multicolumn{2}{c|}{$+O(n^2 \txsize)$} & \multicolumn{4}{c|}{}    \\

     IL w. gossip~(Sec.\ref{sec:incl-list-gossip}) &$+\gossipT$ & $+0$  &$0$  &\multicolumn{2}{c|}{$+O(n \gossipC \txsize)$} & \multicolumn{4}{c|}{}      \\

     IL local ~(Sec.\ref{sec:inclusion-list-local}) &$+0$ & $+0$  &$0$   &\multicolumn{2}{c|}{$+O(n |\inclist|)$} & \multicolumn{4}{c|}{}      \\

     \hline

     Narwhal/HotStuff~\cite{DBLP:conf/eurosys/DanezisKSS22} & $8\delta$ & $4\delta$ & $0$& $O(n^2)$ & $O(n^2)$ & 20& $125K$& $1.8$ &\cite[Fig.6]{DBLP:conf/eurosys/DanezisKSS22}\\
     & & &  & & & 50& $135K$& $1.8$  &\cite[Fig.6]{DBLP:conf/eurosys/DanezisKSS22}\\

     Narwhal/Tusk~\cite{DBLP:conf/eurosys/DanezisKSS22} & $6\delta$ & $4\delta$  & $0$ & $O(n^2)$ & $O(n^2)$ & 20& $160K$& $3.2$ &\cite[Fig.6]{DBLP:conf/eurosys/DanezisKSS22}\\
      & & &  & & & 50& $160K$& $3.2$ &\cite[Fig.6]{DBLP:conf/eurosys/DanezisKSS22}\\

     psync-BullShark~\cite{DBLP:conf/ccs/SpiegelmanGSK22} & $4\delta$ & $4\delta$ & 0 & $O(n^2)$ &$O(n^2)$ & 20& $110K$& $2.5$ &\cite[Fig.2]{DBLP:conf/ccs/SpiegelmanGSK22}\\
      & & &  & & & 50 & $130K$& $2.2$ &\cite[Fig.2]{DBLP:conf/ccs/SpiegelmanGSK22}\\
      Mysticeti~\cite{DBLP:journals/corr/abs-2310-14821}& $3\delta$ & $3\delta$ & 0 & $O(n^2)$ &$O(n^2)$ & 10& $300K$& $<1$ &\cite[Fig.4]{DBLP:journals/corr/abs-2310-14821}\\
      & & &  & & & 50& $100K$& $<1$ &\cite[Fig.4]{DBLP:journals/corr/abs-2310-14821}\\

      \hline
  \end{tabular}
  }
\end{table}

\oa{
\begin{enumerate}

  \item Aequitas, Themis~\cite{DBLP:conf/ccs/KelkarDLJK23}, QOF
  \item Aleph
  \item PaLa, Celestia, Thunderella, Hedera
  \item FnF~\cite{DBLP:conf/sirocco/AvarikiotiHSVWW23}
  \item Economic rewards for producing blocks
  \item Paradigm https://www.paradigm.xyz/2024/02/leaderless-auctions

  \item BBCA~\cite{DBLP:journals/corr/abs-2306-14757}
  \item Mysticeti~\cite{DBLP:journals/corr/abs-2310-14821}
\end{enumerate}
}

\subsection{Single-leader protocols}\label{sec:leader-based}
Tendermint~\cite{DBLP:journals/corr/abs-1807-04938, DBLP:conf/srds/CasonFM0BP21} is a partially synchronous protocol that uses two rounds of voting and an all-to-all communication pattern.
In the happy path, where no faults occur and the network is synchronous, it has a proposal latency of $3 \delta$.
The proposal period is $3 \delta$.
The leader is rotated after every epoch.

HotStuff~\cite{DBLP:conf/podc/YinMRGA19} is a partially synchronous protocol that uses three rounds of voting and an all-to-leader communication pattern.
In the happy path, and assuming an implementation with threshold signatures, it achieves linear communication.
The proposal latency to $7\delta$, as every voting round contains an all-to-leader and a leader-to-all communication step.
A module of the protocol called \emph{pacemaker} is responsible for synchronizing the views of the replicas.
It maintains the current round and, in case of asynchrony or failures, it sends timeout messages to all replicas.
Hence, the worst-case communication complexity is quadratic, because of the all-to-all timeout messages sent by the pacemaker.
The protocol makes no progress while in asynchrony.
Using the technique of \emph{pipelining}, a new proposal can be sent in every round, hence the proposal period is $2\delta$.
HotStuff achieves \emph{optimistic responsiveness}, meaning it can make progress at network speed (i.e., $O(\delta)$) after GST with an honest leader.
The three-round version of HotStuff is also known as DiemBFT~\cite{diemBFT}.

Jolteon~\cite{DBLP:conf/fc/GelashviliKSSX22} is two-round version of HotStuff, achieving a proposal latency of $5\delta$ and linear communication in the happy path.
This is achieved at the cost of a quadratic view-change procedure (after asynchrony or a malicious leader, each replica has to send a message of size $O(n)$ to the next leader).
As the pacemaker is already quadratic, this does not affect the worst-case communication of the protocol.
Ditto~\cite{DBLP:conf/fc/GelashviliKSSX22} is another two-round version of HotStuff, that, like Jolteon, also comes with a quadratic view-change procedure,
but replaces the pacemaker with an asynchronous fallback protocol. This allows the protocol to make progress in case of asynchrony.
The combination of Jolteon over a Narwhal~\cite{DBLP:conf/eurosys/DanezisKSS22} network is known as HotStuff-over-Narwhal and Narwhal/HotStuff.
Finally, HotStuff-2~\cite{DBLP:journals/iacr/MalkhiN23} is also a two-round version of HotStuff, achieving a proposal latency of $5\delta$ in the happy path, at the cost of losing optimistic responsiveness.

MoonShot~\cite{DBLP:journals/corr/abs-2401-01791} builds further on HotStuff-2, adding the idea of
\emph{optimistic proposals}, where a leader can send a new proposal before receiving enough votes for the previous one.
This drops the proposal period to $\delta$ in the happy path at the cost of a quadratic communication complexity and a more complicated protocol logic.
Finally, HotShot\footnote{\url{https://github.com/EspressoSystems/HotShot}} is a Proof-of-Stake version of HotStuff.

\subsection{Parallel-leaders approach}\label{sec:parallel-blocks}

\dotparagraph{Censorship resistance vs. transaction duplication}
Consensus protocols face the following trade-off, stemming from the fact that up to $f$ parties can be malicious:
in order to avoid censorship, a client has to send its transaction to at least $f+1$ replicas, which, depending on the design of the protocol, may lead to request duplication.
As recognized in the literature~\cite{DBLP:journals/corr/abs-1906-05552, DBLP:conf/ccs/MillerXCSS16},
the problem is exacerbated in protocols that feature parallel leaders:
it is not straightforward how to ensure that the leaders, who propose blocks in parallel, do not include the same transactions in them.
In this section we explore how existing protocols handle this trade-off.

\medskip
Several works use the following idea: in each epoch replicas create blocks or `mini-blocks' in parallel
and the protocol outputs a subset of them.
The goal is to limit the leader as much as possible, so it has no other option than produce a correct block, or remain silent.

In HoneyBadger~\cite{DBLP:conf/ccs/MillerXCSS16} replicas collect user transactions in local buffers.
In each epoch they first create and broadcast blocks in parallel,
and then agree on a subset of at least $n-f$ correctly broadcast blocks, which are all output by the protocol.
The authors observe a trade-off between censorship resistance and protocol throughput.
Regarding the censorship resilience vs. transaction duplication dilemma,
the authors propose that replicas include in their block a small number of transactions from their local view, chosen at random,
and that transactions are threshold-encrypted by the clients.
Threshold encryption requires a threshold setup, which either has to be performed by a trusted party or necessitates the
implementation of a Distributed Key Generation (DKG) protocol.
Moreover, it incurs additional computational and communication cost.

DispersedLedger~\cite{DBLP:conf/nsdi/YangPAKT22} builds on this idea, but instead of broadcast it uses a Data Availability (formally, Verifiable Information Dispersal, VID) protocol,
thus allowing replicas to vote for a transaction without locally downloading it.
The protocol guarantees that all blocks proposed by honest replicas will be delivered.
Transaction duplication is not resolved -- in order to make sure it will not be censored,
a client has to send a transaction to $f+1$ replicas, and all of them may include it in their proposed block.

BigDipper~\cite{DBLP:journals/corr/abs-2307-10185} is a system that combines a broadcast, a Data Availability (DA),
and a leader-based consensus protocol to build a censorship resistant leader-based consensus protocol.
The `mini blocks idea' is integrated into their DA protocol:
Replicas collect transactions from clients and batch them into a `mini block'.
The leader receives mini-blocks from replicas,
encodes them appropriately, and disperses the resulting block.
The DA protocol employs 2-dimensional polynomial commitments and Reed-Solomon codes
(similar to state-of-the-art Information Dispersal protocols~\cite{DBLP:conf/aft/NazirkhanovaNT22, cryptoeprint:2024/685})
and BLS signatures, and consists of two rounds of leader-to-all communication.
It achieves the property that, if a client sends a transaction \tx to at least $n-f$ replicas,
then \tx will be included in the next block produced\cite[Table 3]{DBLP:journals/corr/abs-2307-10185}.
The protocol does not handle transaction deduplication, hence fast transaction inclusion comes at the cost
of storing the same transaction multiple time on the DA layer.
The authors show how it can be integrated into HotStuff-2~\cite{DBLP:journals/iacr/MalkhiN23},
but no implementation or benchmark is provided.

Mir-BFT~\cite{DBLP:journals/corr/abs-1906-05552} features parallel leaders, each running a standard leader-based protocol, such as PBFT.
Regarding transaction duplication,
the authors propose partitioning the transaction assignment among the replicas (that is, based on the hash of a transaction, there is one leader responsible for it) and periodically rotating this assignment.
This, however, does not differ much from a single-leader protocol,
as far as censorship is concerned, as, in the worst case, a transaction will be assigned to an honest leader after $f$ such rotations.

Finally, BRAID\footnote{\url{https://ethresear.ch/t/censorship-insurance-markets-for-braid/20288}} is a new proposal for censorship resistance on Ethereum, using the idea of parallel leaders in a rational setting (instead of the Byzantine setting considered by the aforementioned works). The protocol relies on cryptographic primitives such as verifiable delay functions to achieve censorship resistance.

\dotparagraph{Conclusion}
A leader in consensus protocols becomes a temporary point of centralization, and this contributes to censorship.
The aforementioned works aim to completely remove or limit the power of the leader.
On the other hand, employing a leader is a common technique for efficient consensus
(at least, efficient on the so-called `happy path', where the leader is honest and the network is good),
employed by some state-of-the-art protocols~\cite{DBLP:conf/podc/YinMRGA19, DBLP:journals/iacr/MalkhiN23, DBLP:journals/corr/abs-2401-01791,DBLP:journals/corr/abs-1807-04938}.
All aforementioned works achieve censorship resilience at the cost of duplicating transactions, hence wasting computation, communication, and storage.
In \Cref{sec:mini-blocks-construction} we present constructions that achieve censorship resistance using the parallel-blocks idea,
can be employed with minimal modification on existing consensus protocols,
and achieve increasingly better transaction deduplication.

\subsection{DAG-based protocols}\label{sec:dag-based}
The so-called `DAG-based' protocols observe that the separation of data dissemination and ordering logic improves the efficiency
of consensus protocols.
Assuming that clients send transactions to `enough' (explained in the next paragraph) replicas,
DAG-based protocols achieve short-term censorship resistance by construction, as blocks are created by all replicas in parallel.
This comes, unavoidably, with transaction duplication.

In the asynchronous Narwhal/Tusk~\cite{DBLP:conf/eurosys/DanezisKSS22} the blocks of up to $f$ honest but slow replicas can be arbitrarily delayed (even gargabe-collected, hence never delivered).
With up to $f$ replicas being malicious, in order to achieve short-term censorship resistance transactions have to be sent to $2f+1$ replicas.
The partially synchronous BullShark~\cite{DBLP:conf/ccs/SpiegelmanGSK22,DBLP:journals/corr/abs-2209-05633} protocol guarantees that, after GST, the blocks of all honest replicas will become delivered.
Hence, assuming being in a synchronous period, clients can send transactions to $f+1$ replicas.

Mysticeti~\cite{DBLP:journals/corr/abs-2310-14821} achieves very low latency if there are no Byzantine faults, which the authors argue is the most common case in practice.
The improvement comes mainly from using an uncertified DAG, where blocks are multicast and not broadcast.
This allows blocks to be sent and committed within three network trips, hitting the lower bound for consensus. Multicasting also allows validators to equivocate, by sending two different blocks. If that happens, either only one of them will be committed, or none will be committed for that epoch. Malicious behavior like this and asynchrony lead to a less efficient fallback `indirect decision rule'.
Mysticeti also provides built-in support for fast-path transactions, that is transactions that do not need to be totally ordered (see \Cref{sec:consensusless}).

Regarding practical efficiency, as shown in \Cref{tab:comparison},
Narwhal/Tusk achieves the highest throughput (160K with 50 replicas), but also the highest latency.
The partially synchronous version of BullShark~\cite{DBLP:conf/ccs/SpiegelmanGSK22,DBLP:journals/corr/abs-2209-05633},
maintains comparable throughput (130K with 50 replicas) and a better latency, but still over 2 sec.
Narwhal/HotStuff~\cite{DBLP:conf/eurosys/DanezisKSS22} achieves similar throughput (135K with 50 replicas) and the lowest latency, approx. 1.8 seconds.
Mysticeti~\cite{DBLP:journals/corr/abs-2310-14821} achieves a throughput of around 300K tps in a deployment with 10 replicas, and 100K tps in a deployment with 50 replicas, while maintaining sub-second latency.

A significant advantage of DAG-based, compared to leader-based, protocols is their better resilience to crash faults.
This is because they do not employ view-changes.
For example, in benchmarks with ten replicas, BullShark achieves a throughput of $70K$ tps when three replicas crash,
while its latency becomes approx. 6 seconds~\cite[Fig.~4]{DBLP:conf/ccs/SpiegelmanGSK22}.
In the same experiment, Narwhal/HotStuff~\cite{DBLP:conf/eurosys/DanezisKSS22} also achieves a throughput of $70K$ tps with a latency or approx. 10 seconds~\cite[Fig.~8]{DBLP:conf/eurosys/DanezisKSS22},
while HotStuff achieves a throughput of $10K$ tps with approx. 14 seconds latency~\cite[Fig.~8]{DBLP:conf/eurosys/DanezisKSS22}.

\dotparagraph{Conclusion}
DAG-based protocols outperform leader-based protocols in terms of throughput,
while exhibiting comparable latency, in the best case approx. 2 seconds.
In order to achieve sub-second latency in production systems they have been combined with consensusless protocols~\cite{DBLP:journals/corr/abs-2310-18042,DBLP:conf/aft/BaudetDS20}.

\subsection{Consensusless protocols}\label{sec:consensusless}
Recent literature has recognized the redundancy of consensus for implementing asset-transfer systems~\cite{DBLP:journals/dc/GuerraouiKMPS22}.
Such schemes have been described in theory~\cite{DBLP:journals/corr/abs-1909-10926, DBLP:conf/dsn/CollinsGKKMPPST20} and deployed in the real world~\cite{DBLP:conf/aft/BaudetDS20}.

The insight that is total order is not required in the case that each account is controlled by one client.
Instead, it is sufficient to guarantee that cheating clients cannot equivocate, that is, send different transactions to different replicas.
This property is guaranteed by broadcast protocols.
These protocols have a similar architecture: the broadcast of transactions is initiated by clients, who either drive the whole broadcast instance~\cite{DBLP:conf/aft/BaudetDS20} or outsource it to trusted replicas~\cite{DBLP:conf/dsn/CollinsGKKMPPST20}.
Therefore, a cheating client might lose liveness~\cite{DBLP:conf/aft/BaudetDS20,DBLP:journals/corr/abs-1909-10926}, but equivocating is not possible.

Specifically, in FastPay~\cite{DBLP:conf/aft/BaudetDS20} a client sends its transaction (a payment to some recipient) to all replicas, waits for $n-f$ signatures on it,
and forms a certificate with them. The certificate is enough for the sender and the recipient to consider the payment finalized,
because it proves that no conflicting transaction can ever be accepted by the replicas.
The replicas update the balance of the sender and the recipient when they receive the certificate from the client.
A necessary component in the construction is a sequence number maintained by each client: transactions submitted by a client must have
consecutive sequence numbers, and no transaction may be pending (a transaction is \emph{pending} when a replica has signed it but not received the certificate for it) when the client submits the next one.
The sequence number is exactly what provides safety for payments: clients cannot equivocate (e.g., double-spend) because they can submit at most one
transaction per sequence number, and all transactions submitted by a client are ordered.
Malicious client, trying to send conflicting transactions for a sequence number, may lose liveness by not being able to form a certificate for any of them.

Astro~\cite{DBLP:conf/dsn/CollinsGKKMPPST20} generalizes the sequence number to an \emph{xlog}, an append-only log that contains all outgoing payments from each account,
maintained by the single owner of that account.
Only the account owner can broadcast updates to it xlog, hence Astro guarantees total order within each xlog and achieves safety for payments.
ABC~\cite{DBLP:journals/corr/abs-1909-10926} is similarly based on reliable broadcast~\cite{DBLP:journals/iandc/Bracha87}. In addition to transactions, replicas can
broadcast \emph{votes} for transactions they have seen, and the votes are weighted by the replica's stake.
This enables the system to also work in permissionless settings.

FastPay~\cite{DBLP:conf/aft/BaudetDS20} reaches throughput of $140K$ tps with a latency of approx. 200 ms in a WAN deployment with four replicas.
Astro~\cite{DBLP:conf/dsn/CollinsGKKMPPST20} achieves a throughput of $5K$ tps with latency of approx. 200 ms~\cite[Fig.4, Astro II]{DBLP:conf/dsn/CollinsGKKMPPST20} in a WAN deployment with 100 replicas.
ABC~\cite{DBLP:journals/corr/abs-1909-10926} provides no implementation or benchmarks.

\dotparagraph{Conclusion}
To the best of our knowledge, the only consensusless system that has been used in production and supports both payments and arbitrary objects (data declared in smart contracts)
is FastPay in the Sui Blockchain~\cite{DBLP:journals/corr/abs-2310-18042}.
However, that same work observes that consensusless protocols cannot offer checkpointing
and are prone to losing liveness even for honest clients, due, for example, to clients' misconfigurations.
For this reason, the consensusless protocol is combined with a DAG-based consensus protocol~\cite{DBLP:journals/corr/abs-2310-18042}.
Transactions that do not need total order can be executed as long as the client broadcasts them,
but \emph{all} transactions eventually go through the consensus protocol.
Hence, the system offers significantly better latency, but a consensus protocol is still required, and it must be able to handle the total workload of the system.
Since different replicas hold different state at any moment in a consensusless protocol,
the combination with a consensus protocol also allows light clients to deterministically read a consistent state from the system.

Moreover, all these protocols are tailor-made for payment systems and cannot be used for general distributed applications.
They employ, directly or indirectly, sequence numbers in order to achieve total-order for transactions sent by each client,
a property that is required~\cite{DBLP:journals/dc/GuerraouiKMPS22} for payment systems but not for other use cases, such as auctions.

\subsection{Separating block builders and proposers}
Chop chop~\cite{DBLP:journals/corr/abs-2304-07081} introduces a new layer, called the \emph{brokers}, between clients and replicas running a consensus protocol.
Brokers are responsible for building blocks of transactions in a way that minimizes the transaction metadata (such as client signatures) in a block. This allows blocks to contain a larger number of transactions
resulting in a system with higher throughput, compared to the underlying consensus protocol.
On the other hand, brokers engage in interactive protocols with the clients and the replicas, hence increasing the time needed for a transaction to get committed.
The system can support multiple brokers, but each of them runs a non-distributed protocol. Hence, fairness and censorship resistance are not achieved.
Encrypting client transactions would not be enough -- the brokers need to know the client behind each transaction because they engage in an interactive multi-signature protocol, hence they can censor specific clients.

\section{Leader-based protocols with Inclusion Lists}\label{sec:mini-blocks-construction}
In this section we present an approach that can be combined with any leader-based protocol (such as Tendermint\cite{DBLP:journals/corr/abs-1807-04938} or HotStuff~\cite{DBLP:conf/podc/YinMRGA19}).
It changes the way a proposal is created and voted for.
We assume an underlying protocol that proceeds in \emph{epochs} and each epoch has a unique \emph{leader} that creates a \emph{block}.
We assume a partially synchronous network. A high-level analysis of each construction is also provided.

These protocols are based on the idea of having each replica create an \emph{Inclusion List (IL)}, a list of transactions,
and then restricting the leader to use \emph{only} the ILs when creating a block.
\Cref{sec:inclusion-lists-base} presents the base protocol, while the following sections present optimizations.

\subsection{Leader-based consensus with Inclusion Lists}\label{sec:inclusion-lists-base}

\dotparagraph{The protocol}
Clients submit transactions to replicas. On every epoch each replica creates an IL,
signs it and sends it to the leader of that epoch.
The leader waits for $n - f$ ILs from distinct replicas and creates a block that contains \emph{only} the
$n-f$ ILs and no other transactions.
Upon receiving the proposal, each replica sends a vote if it considers the block valid.
The block is valid if (in addition to the conditions of the underlying leader-based protocol) it contains at least $n - f$ inclusion
lists, each signed by a different replica.
Note that role of the leader now only consists in choosing which $n - f$ (or more) inclusion lists will be used in the new block.

\dotparagraph{Properties}
The protocol achieves the same safety and liveness properties as the underlying leader-based protocol,
and the additional \emph{short-term censorship resistance} property.
Note that, as in the underlying protocol, liveness can be attacked by malicious leaders (e.g., by remaining silent and not producing any block), but selective censorship is not possible.
We present arguments to incentivize leaders to produce blocks on \cref{sec:economic-args}.


Censorship resistance comes from the fact that the leader can only ignore up to $f$ inclusion lists.
If it ignores more, no honest replica will vote for the proposal.
Hence, the client needs to ensure that $f+1$ honest replicas have received its transaction.
This can be achieved by sending it to $2f + 1$ replicas.

\dotparagraph{Special cases}
\begin{itemize}
    \item
    It can be the case that not all transactions in the $n-f$ inclusion lists fit in the next block. To maintain fairness, a deterministic rule is needed for the leader to choose which transactions to add.
    One option is to have the leader add transactions by frequency of appearance in the inclusion lists.
    A second is to require that transactions are ordered in the inclusion lists and the leader selects the first $x$ transactions from each inclusion list,
    such that $x$ is as large as possible given the block size.
    \item
    Contradictory transactions may exist in the inclusion lists, such as two transactions from a client who can only pay the fees for one of them.
    We can again break the tie in a deterministic way, for example by keeping the transaction from the inclusion list of the replica with the lowest identifier.
\end{itemize}

\dotparagraph{Analysis}
For an overview of the construction we refer to \Cref{tab:comparison}.
Our modification can be implemented by having every replica send its inclusion list together with the last vote message of the previous epoch.
For example, if implemented on Tendermint~\cite{DBLP:journals/corr/abs-1807-04938}, the inclusion lists can be sent using ABCI++, piggybacked on vote messages.
The proposal latency and proposal period hence remain unchanged.
Assuming the leader does not remain silent and none of the special conditions explained above applies, an honest client's transaction will be included in the next block, that is, \emph{max tx censorship} is $0$.
However, similar to the protocols presented in \Cref{sec:parallel-blocks}, the construction leads to transaction duplication,
as a transaction may appear in multiple inclusion lists.
We denote this in \Cref{tab:comparison} as an $O(n^2 \cdot |\inclist|)$
additional communication cost, as the leader has to include to its proposal $O(n)$ inclusion lists of average size $|L|$.
We present mitigations in the following sections.

\subsection{Using a Data Availability layer}\label{sec:incl-list-da}
In this version, the inclusion lists contain \emph{references} to transactions.
The full transactions are submitted by the client to a Data Availability (DA) layer.
We abstract the DA layer as follows.

\begin{definition}[Data Availability (DA) scheme~\cite{DBLP:conf/aft/NazirkhanovaNT22}]
    A \emph{DA} scheme is run among \emph{clients} and \emph{storage servers}.
    It exposes the following algorithms, which are initiated by a client and by the client and all storage nodes.
    \begin{itemize}
        \item \opdisperse{\tx} \returns $P$\footnote{We abstract the commitment $C$ from~\cite{DBLP:conf/aft/NazirkhanovaNT22} inside $P$.}:
        It takes as input a transaction \tx and returns a \emph{certificate of retrievability} $P$.
        \item \opretrieve{$P$} \returns $\tx$:
        It takes as input a certificate of retrievability $P$ and returns a transaction \tx or $\bot$.
    \end{itemize}
    If an honest client invokes \opdisperse{\tx}, then it will obtain a certificate of retrievability $P$,
    such that, if an honest (and possibly different) client invokes \opretrieve{$P$}, then the second client will obtain \tx.
    Moreover, all calls to \opretrieve{} return the same value to all honest clients, except with negligible probability,
    even if the client that initiated \opdisperse{} was malicious (in which case \opretrieve{} may return $\bot$).
\end{definition}

\dotparagraph{The protocol}
The client firsts submits transaction \tx to the DA layer. Once it obtains the certificate of availability $P$, it sends it to all replicas.
Upon receiving $P$, if $\opretrieve{P} \neq \bot$ then a replica appends $P$ to its IL, which is forwards to the leader.
The leader waits for $n-f$ valid ILs, where an IL is valid if,
for all certificates of availability $P$ it contains, it holds that $\opretrieve{P} \neq \bot$.
The leader creates a block that contains all transactions retrieved from the $n-f$ valid ILs.
The leader sends a proposal with the new block and the $n-f$ signed ILs to all replicas.
The proposal is valid if it contains $n-f$ ILs and the block contains all corresponding transactions.

\dotparagraph{Analysis}
Let \dadisp denote the average time of \opdisperse{} and \daretr that of \opretrieve{}. This construction effectively increases \emph{proposal latency} by $\dadisp + \daretr$,
because a client has to disperse \tx and the leader checks that it can be retrieved. If the leader produces some block, then an honest clinet's transaction will be included in it, that is, \emph{max tx censorship} is 0. Finally, \emph{communication complexity} increases by a factor of \dacom, depending on the implementation of the DA layer.
We show these on \Cref{tab:comparison}.

\dotparagraph{Advantages and drawbacks}
The inclusion list can now contain pointers to transactions, while the actual payload exists only in the DA layer.
On the other hand, the DA layer adds latency to the protocol.

\dotparagraph{Further optimizations}
\begin{itemize}[left=0pt,itemindent=0pt]
    \item In order to further reduce the output size, the leader can write the certificates of availability -- instead of
    the corresponding transactions -- in the block. This comes at the cost or requiring clients to query the DA layer and retrieve it.
    \item We can allow clients to submit invalid certificates of availability, i.e., $P$ for which $\opretrieve{P} = \bot$.
    This works because, by the properties of the DA scheme, clients that read the output of our protocol will agree on the
    output of $\opretrieve{P}$.
    The drawback of this is that the output can contain garbage transactions.
\end{itemize}

We remark that the proposed construction is similar to BigDipper~\cite{DBLP:journals/corr/abs-2307-10185}, with the following differences.
First, the DA layer is here decoupled from the consensus layer, and it is the client's responsibility to disperse the transaction.
Second, our protocol achieves transaction deduplication, as the leader includes each transaction only once in the proposed block.

\subsection{Using reliable broadcast}\label{sec:incl-list-bcast}
Instead of using a separate Data Availability layer, in this section we have the client broadcast \tx to the replicas.

\dotparagraph{The protocol}
The client sends a transaction \tx using a version of reliable broadcast~\cite{DBLP:journals/iandc/Bracha87}.
The broadcast algorithm consists of three communication steps. On the first, the client sends \tx  to all  replicas. The other two consist of all-to-all communication among the replicas.
When a replica delivers \tx, it adds to its inclusion list the hash of \tx.
By properties of reliable broadcast, if an honest replica delivers \tx, then all honest will eventually deliver \tx.
Hence, for every IL of an honest replica, the leader will eventually deliver all included transactions.
The leader includes in the new block the first $n-f$ ILs whose transactions are delivered in the broadcast layer.

\dotparagraph{Analysis}
As shown on \Cref{tab:comparison}, when implemented on top of a leader-based protocol, this construction effectively increases \emph{proposal latency} by $2\delta$, because reliable broadcast requires two additional communication rounds.
The \emph{proposal period} remains unchanged, while \emph{communication complexity} increases by a factor of $O(n^2 \cdot s)$,
where \txsize is the average transaction size.

\dotparagraph{Advantages and drawbacks}
The inclusion lists, appended to the new block, can now contain hashes of transactions, and not the transactions themselves, thus reducing the size of the block.
On the other hand, the protocol adds two all-to-all communication rounds to the underlying consensus protocol.
\oa{Is there a way to eliminate the second all-to-all round (ready messages)? Maybe we can make the inclusion list function as the ready message. But then the inclusion list would have to contain the whole transaction.}


Notice that, different to DAG-based approaches, broadcasts in this construction are initiated by the clients,
and they are performed on transaction and not block level.
Hence, we can avoid transaction duplication.

\subsection{Using a gossip layer}\label{sec:incl-list-gossip}
Instead of a broadcast primitive, we can use a gossip layer to make transactions available to all parties.

\oa{todo: Model the gossip layer.}

\dotparagraph{The protocol}
The only difference from the previous section is that replicas do not broadcast the transactions received from clients, but they gossip them to each other.
The ILs contain again pointers to transactions.
Since there are at least $n-f$ honest parties, and assuming the gossip layer has been instantiated correctly to allow propagation of transactions to all replicas,
the leader will eventually receive $n-f$ ILs, such that it has received the corresponding transactions via the gossip layer.

\dotparagraph{Analysis}
Let \gossipT denote the average propagation time and \gossipC the number of replicas each replica connects to in the gossip-layer implementation, and \txsize the average transaction size.
As shown on \Cref{tab:comparison}, when implemented on top of a leader-based protocol, this increases \emph{proposal latency} by \gossipT and \emph{communication complexity} by a factor of $O(n \cdot \gossipC \cdot s)$.

\dotparagraph{Advantages and drawbacks}
Compared to the broadcast based, this solution does not require two additional
rounds of communication for every transaction.
Moreover, replicas need to maintain fewer network connections, as there is no all-to-all communication.

\subsection{A protocol without writing the Inclusion Lists on the block}\label{sec:inclusion-list-local}
We now present a modification to the protocol in~\Cref{sec:inclusion-lists-base} which does not require the leader to
append the $2f+1$ used ILs in the proposal message.

\dotparagraph{The protocol}
Similar to~\Cref{sec:inclusion-lists-base} each replica sends its IL to the leader. The leader chooses $n-f$ and creates a block with their transactions. In the proposal message the leader
includes the \emph{lists-used} field, a list with the identifiers of the replicas whose ILs it used.
Replicas vote for a proposal only if contains a \emph{lists-used} field  of size at least $2f+1$. Additionally, a replica whose identifier is in
the \emph{lists-used} field verifies whether its IL is indeed in the transactions of the new block.

On \Cref{tab:comparison} we summarize the trade-offs of this solution. Compared to the protocol in \Cref{sec:inclusion-lists-base},
this only incurs an additional communication cost of $O(n \cdot |L|)$, where $|L|$ is the average size of an inlcusion list,
as each replica sends one IL to the leader.

\dotparagraph{Design choices and correctness}
Note that the leader must send the proposal and consider the votes from all the replicas.
If it sent it only to the $2f+1$ whose IL it used, or counted the votes only from them, then a single malicious replica among these $2f+1$ would be able to harm liveness.
In other words, the $f$ replicas whose IL was not used by the leader have to vote for the proposal,
without being able to verify whether the leader actually included all the transactions from the \emph{lists-used} field.
Observe that these might be honest replicas.
Moreover, $f$ votes can come from malicious replicas,
hence the leader needs only one vote from an honest replica in the \emph{lists-used} field.
This means that the leader only needs to actually use \emph{one} IL sent by an honest replica, when it claims to have used $2f+1$.

\dotparagraph{Censorship resistance}
In this protocol the leader can ignore up to $2f$ inclusion lists from honest replicas.
Hence, the client needs to ensure that $2f+1$ honest replicas have received its transaction.
This can be achieved by sending it to all replicas.
We comment on \Cref{sec:economic-args} on how this translates to worse censorship resistance, compared to the rest of the protocols
in this section.

\subsection{Related works}
  Protocols such as BigDipper~\cite{DBLP:journals/corr/abs-2307-10185}, DispersedLedger~\cite{DBLP:conf/nsdi/YangPAKT22} can be seen as implementations of the inclusion lists approach.
  The notion of inclusion lists also appears on Ethereum-focused research\footnote{\url{https://eips.ethereum.org/EIPS/eip-7547}}.

\section{Economic arguments}\label{sec:economic-args}

\dotparagraph{The economic-censoring model}
When reasoning about the censorship resistance of a protocol, we work with two models. The first is the \emph{honest-malicious} setting,
where \emph{honest} replicas follow the protocol, and hence do not censor any transactions they have received,
and \emph{malicious} replicas can behave arbitrarily.
The second is the \emph{economic-censoring} model, which is the same as  the honest-malicious, but replicas (both honest and malicious) have one additional choice: for each received client transaction, they decide  whether to ignore it or not.
They base this choice on economic criteria, which we abstract in the notion of a \emph{bribery}. A bribery for a transaction \tx is an amount of money greater than the reward a replica would get for including that transaction.
If a replica is bribed to censor \tx, then it will censor it, and if it is not bribed, then it will not censor it.

\dotparagraph{Censorship resistance of the protocol in Sections~\ref{sec:inclusion-lists-base}--\ref{sec:incl-list-gossip}}
In order for the adversary to delay the inclusion of a transaction for one epoch,
it would have to bribe the leader of that epoch and $2f$ replicas.

\dotparagraph{Censorship resistance of the protocol in~\Cref{sec:inclusion-list-local} }
In order for the adversary to delay the inclusion of a transaction for one epoch,
it would have to bribe the leader of that epoch and $f$ replicas.

\section{Conclusion and recommendations}\label{sec:conclusion}
In this report we evaluated consensus protocols with regard to their efficiency and short-term censorship resistance.
For applications that rely on these properties (such as decentralized auctions, decentralized sequencers, data-feed applications, etc.),
we make the following recommendations:

\dotparagraph{Parallel-leader protocols}
Parallel-leader protocols (\Cref{sec:parallel-blocks}) satisfy the definition of short-term censorship resistance by construction.
Yet, they pay the price of transaction duplication, are relatively inefficient,
and feature no production implementation.

\dotparagraph{DAG-based}
DAG-based protocols (\Cref{sec:dag-based}) also achieve short-term censorship resistance, but with duplicate (specifically, up to $2f+1$ copies) transactions in the output of the protocol.
They reach comparatively high throughput, but suffer from high latency.
This is the reason why in production they have been combined with consensusless protocols~\cite{DBLP:journals/corr/abs-2310-18042}, resulting in sub-second latency.
Applications can use (or build on existing codebases of)
Narwhal/BullShark~\cite{DBLP:conf/ccs/SpiegelmanGSK22}, aiming for latency in the order of 2 seconds,
or Mysticeti~\cite{DBLP:journals/corr/abs-2310-14821}, aiming for sub-second latency.
We observe, however, that, if we do not count for duplicate transactions in the output of the protocol, the throughput of DAG-based protocols is not expected to differ much from single-leader protocols (see \Cref{tab:comparison}).

\dotparagraph{Leader-based protocols with a censorship-resistance add-on component}
In \Cref{sec:leader-based} we reviewed and compared existing leader-based protocols,
and then proposed censorship-resistance solutions that can be implemented
on top of them (\Cref{sec:mini-blocks-construction}).
A leader-based protocol with an Inclusion List add-on component (\Cref{sec:mini-blocks-construction})
would achieve the desired definition of short-term censorship resistance.
The constructions in Sections~\ref{sec:incl-list-da}--\ref{sec:inclusion-list-local}
avoid transaction duplication.
The one in \Cref{sec:inclusion-list-local} does not require additional communication rounds.
As shown in \Cref{sec:economic-args}, the constructions in Sections~\ref{sec:inclusion-lists-base}--\ref{sec:incl-list-gossip}
achieve, in a rational setting, better censorship resistance.
Of advantage here is the existence of production implementations, in particular of Tendermint, but also of HotStuff.
The drawback with this approach is the low throughput and high latency of single-leader protocols,
as well as the performance deterioration in presence of crash faults.

\dotparagraph{The pod protocol}
Finally, applications that do not require total ordering of transactions, but instead prioritize fast transaction confirmation and censorship resistance, can use the protocol of pod~\cite{DBLP:journals/corr/abs-2501-14931}. Pod achieves the physically optimal latency of one round trip for confirmation, and by design avoids parties with `special' power, such as leaders, who could censor transactions.

\section{Existing implementations}\label{sec:implementations}

In this section we provide references to implementations of the protocols mentioned throughout the report.

The 3-round version of HotStuff~\cite{DBLP:conf/podc/YinMRGA19} (see HotStuff/DiemBFT in \Cref{tab:comparison} and \Cref{sec:leader-based}) has been implemented\footnote{\url{https://github.com/asonnino/hotstuff/tree/3-chain}} in Rust, but the authors state it is not production ready.
The same protocol is available\footnote{\url{https://github.com/diem/diem/tree/latest/consensus}} as part of Diem's codebase, again in Rust.
A modular, academic implementation\footnote{\url{https://github.com/relab/hotstuff/tree/master/consensus/chainedhotstuff}} exists in Go,
a prototype implementation\footnote{\url{https://github.com/gitferry/bamboo/tree/master/hotstuff}} as part of the Bamboo~\cite{DBLP:conf/icdcs/GaiFNFBD21} framework also exists in Go,
while the academic prototype\footnote{\url{https://github.com/hot-stuff/libhotstuff}} for the original paper was written in C++.

The 2-round version of HotStuff~\cite{DBLP:conf/fc/GelashviliKSSX22} (see Jolteon in \Cref{tab:comparison} and \Cref{sec:leader-based}) has been implemented\footnote{\url{https://github.com/asonnino/hotstuff/}}\footnote{\url{https://github.com/relab/hotstuff/tree/master/consensus/fasthotstuff}}\footnote{\url{https://github.com/gitferry/bamboo/tree/master/fasthostuff}} in some of the aforementioned repositories.
Ditto, the 2-round version of HotStuff with an asynchronous fallback protocol, has also been implemented\footnote{\url{https://github.com/danielxiangzl/Ditto}} in Rust.
A prototype implementation\footnote{\url{https://github.com/facebookresearch/narwhal/tree/narwhal-hs}} of HotStuff-over-Narwhal is also available.

Regarding DAG-based protocols, Narwhal/Tusk~\cite{DBLP:conf/eurosys/DanezisKSS22}, that is, the asynchronous Tusk consensus protocol over Narwhal, is available\footnote{\url{https://github.com/asonnino/narwhal}} in Rust.
Narwhal/Bullshark, that is, the partially-synchronous BullShark consensus protocol over Narwhal, has also been implemented\footnote{\url{https://github.com/asonnino/narwhal/tree/bullshark}} in Rust.
A prototype implementation of Mysticeti~\cite{DBLP:journals/corr/abs-2310-14821} is also available\footnote{\url{https://github.com/MystenLabs/mysticeti}}.
Mysten labs provides implementations of Narwhahl\footnote{\url{https://github.com/MystenLabs/sui/tree/main/narwhal}}, as well as Narwhal/Tusk and Narwhal/Bullshark~\footnote{\url{https://github.com/MystenLabs/narwhal}}.

The Tendermint consensus algorithm~\cite{DBLP:journals/corr/abs-1807-04938}, also known as CometBFT, has been implemented\footnote{\url{https://github.com/cometbft/cometbft}} in Go and in Rust\footnote{\url{https://github.com/informalsystems/tendermint-rs}}.
FastPay has been implemented\footnote{\url{https://github.com/novifinancial/fastpay}} by Facebook in Rust.

\vfill
\pagebreak
\iflncs
\thispagestyle{plain}
\fi

\ifccs
  \bibliographystyle{ACM-Reference-Format}
\else
  \bibliographystyle{abbrv}
\fi

\pdfbookmark[section]{References}{references}
\bibliography{biblio,crypto,pubs,dblpbibtex}

\section*{About Common Prefix}

Common Prefix is a blockchain research, development, and consulting
company consisting of a small number of scientists and engineers
specializing in many aspects of blockchain science.
We work with industry partners who are looking to advance the state-of-the-art
in our field to help them analyze and design simple but rigorous protocols
from first principles, with provable security in mind.

Our consulting and audits pertain to theoretical cryptographic
protocol analyses as well as the pragmatic auditing of implementations
in both core consensus technologies and application layer smart contracts.

\begin{figure}
    \center
    \includegraphics[width=0.7\columnwidth]{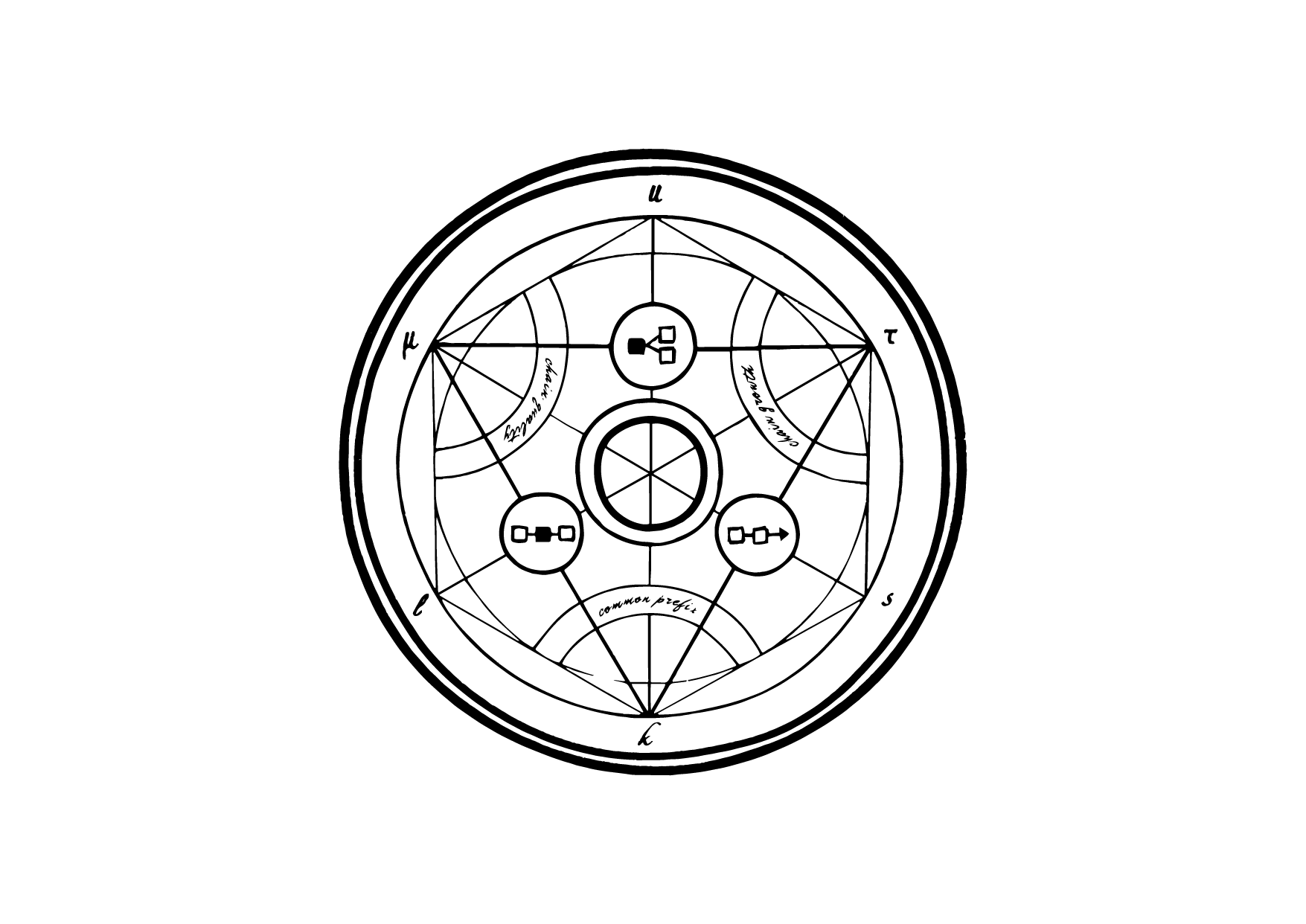}
    \label{fig:cp_logo}
\end{figure}

\end{document}